\documentclass[journal]{IEEEtran}

\usepackage{amsmath,amssymb,amsfonts}
\usepackage{algorithmic}
\usepackage{graphicx}
\usepackage{textcomp}
\def\BibTeX{{\rm B\kern-.05em{\sc i\kern-.025em b}\kern-.08em
    T\kern-.1667em\lower.7ex\hbox{E}\kern-.125emX}}

\DeclareGraphicsExtensions{.pdf,.jpeg,.png}

\usepackage{xcolor}
\usepackage{color}
\usepackage{amsfonts}
\usepackage{amssymb}
\usepackage{array}
\usepackage{booktabs}
\usepackage{multirow}
\usepackage{hyperref}
\usepackage[nolist]{acronym}
\usepackage[normalem]{ulem}
\usepackage{tabularx,colortbl}

\usepackage{cite}
\usepackage{url}
\usepackage{soul}
\usepackage{tikz}
\usepackage{pgfplots}

\newcommand{\J}{\ensuremath{\mathrm{j}}}        
\newcommand{\Quot}[1]{``{#1}"}                  

\providecommand{\V}[1]{\boldsymbol{#1}}         
\providecommand{\M}[1]{\mathbf{#1}}             
\providecommand{\T}[1]{\mathrm{#1}}             
\providecommand{\UV}[1]{\V{\hat{#1}}}           

\providecommand{\basisFcn}{\V{\psi}}

\providecommand{\Ivec}{\ensuremath{\M{I}}}
\providecommand{\Vvec}{\ensuremath{\M{V}}}

\providecommand{\Xin}{\ensuremath{X_\mathrm{A}}}

\providecommand{\herm}{\mathrm{H}}

\newcommand{\ie}{\textit{i.e.}{}}
\newcommand{\eg}{\textit{e.g.}{}}

\providecommand{\Jsurf}{\V{J}}
\providecommand{\BFmat}{\V{\Psi}}
\providecommand{\Zmat}{\M{Z}}
\providecommand{\Rmat}{\M{R}}
\providecommand{\Xmat}{\M{X}}

\providecommand{\surfres}{R_\T{s}}
\providecommand{\radeff}{\eta}
\providecommand{\dissipfact}{\delta}
\providecommand{\Prad}{P_\T{rad}}
\providecommand{\Preact}{P_\T{react}}
\providecommand{\Plost}{P_\T{loss}}
\providecommand{\Qtun}{Q_\mathrm{T}}

\newcommand{\BE}{\begin{equation}}
\newcommand{\EE}{\end{equation}}
\newcommand{\BEn}{\begin{equation*}}
\newcommand{\EEn}{\end{equation*}}
\newcommand{\BF}{\begin{figure}\centering}
\newcommand{\EF}{\end{figure}}
\newcommand{\BT}{\begin{table}\centering}
\newcommand{\ET}{\end{table}}

\newcommand\figwidth{8.9} 

\hypersetup{
	colorlinks=true, 
	linkcolor=blue!70!black, 
	citecolor=red!70!black, 
	filecolor=blue!70!black, 
	urlcolor=blue!70!black, 
}

\newacro{MoM}[MoM]{method of moments}
\newacro{MOO}[MOO]{multiobjective optimization}
\newacro{CM}[CM]{characteristic mode}
\newacro{PEC}[PEC]{perfect electric conductor}
\newacro{PMC}[PMC]{perfect magnetic conductor}
\newacro{EP}[EP]{eigenvalue problem}
\newacro{GEP}[GEP]{generalized eigenvalue problem}
\newacro{EFIE}[EFIE]{electric field integral equation}
\newacro{SVD}[SVD]{singular value decomposition}
\newacro{RWG}[RWG]{Rao-Wilton-Glisson}
\newacro{EM}[EM]{electromagnetic}
\newacro{dof}[d-o-f]{\mbox{degrees-of-freedom}}

\begin{document}

\title{Radiation Efficiency Cost of Resonance Tuning}
\author{Lukas~Jelinek, Kurt~Schab,~\IEEEmembership{Member,~IEEE}, and Miloslav~Capek,~\IEEEmembership{Senior~Member,~IEEE}
\thanks{Manuscript received \today; revised \today.}
\thanks{This work was supported by the Czech Science Foundation under Project 15-10280Y. Kurt Schab was supported by a grant from the Intelligence Community Postdoctoral Research Fellowship Program.  All statements of fact, opinion, or analysis expressed are those of the author and do not reflect the official positions or views of the Intelligence Community or any other U.S. Government agency.  Nothing in the contents should be construed as asserting or implying U.S. Government authentication of information or Intelligence Community endorsement of the author’s views.}
\thanks{L.~Jelinek and M.~Capek are with the Department of Electromagnetic Field, Faculty of Electrical Engineering, Czech Technical University in Prague, Technicka~2, 16627, Prague, Czech Republic
	(e-mail: \mbox{lukas.jelinek@fel.cvut.cz}, \mbox{miloslav.capek@fel.cvut.cz}).}
\thanks{K.~Schab is with the Department of Electrical Engineering, Santa Clara University, Santa Clara, CA, USA (e-mail: kschab@scu.edu).}
}

\markboth{Journal of \LaTeX\ Class Files,~Vol.~XX, No.~XX, August~2018}
{Jelinek \MakeLowercase{\textit{et al.}}: Efficiency cost of tuning}
 \maketitle

\begin{abstract}
Existing optimization methods are used to calculate the upper-bounds on radiation efficiency with and without the constraint on self-resonance.  These bounds are used for the design and assessment of small electric-dipole-type antennas.  We demonstrate that the assumption of lossless, lumped, external tuning skews the true nature of radiation efficiency bounds when practical material characteristics are used in the tuning network.  A major result is that, when realistic (\eg{}, finite conductivity) materials are used, small antenna systems exhibit dissipation factors which scale as $(ka)^{-4}$, rather than $(ka)^{-2}$ as previously predicted under the assumption of lossless external tuning.
\end{abstract}

\section{Introduction}
\label{Intro}

\IEEEPARstart{R}{adiation} efficiency is a parameter of paramount importance for electrically small radiators since it significantly limits antenna performance~\cite{Fujimoto_Morishita_ModernSmallAntennas}. Techniques for maximizing radiation efficiency~\cite{Smith_1972_TAP,Smith_1977_TAP} or attempts to set physical bounds~\cite{Fujita_max_gain_APS_2013,2013_Karlsson_PIER,Fujita_max_gain_IEICE_TE_2015,2018_Shahpari_TAP,Pfeiffer_FundamentalEfficiencyLimtisForESA,2017_Losenicky_Comment_Pfeiffer,Thal2018_RadiationEfficiencyLimits} on this parameter therefore naturally accompany developments in antenna technology.

In most cases, it is desirable to design a small antenna to be tuned (\ie{}, resonant) at a specified frequency.  This can be accomplished either by designing the antenna itself to be self-resonant or through the use of an external tuning network. Although the high cost of resonance tuning in radiation efficiency was recognized long ago \cite{Smith_1977_TAP}, many recent works assume that resonance tuning can be done in a lossless manner using external networks \cite{Fujita_max_gain_APS_2013,2013_Karlsson_PIER,Fujita_max_gain_IEICE_TE_2015,2018_Shahpari_TAP}. In fact, careful review reveals that this is a common assumption in many standard textbooks \cite{2012_Stutzman_Antenna_Theory,VolakisChenFujimoto_SmallAntennas,Fujimoto_Morishita_ModernSmallAntennas} as well. This assumption, however, leads to physical bounds which are unachievable by realistically tuned antenna designs~\cite{BestHanna_AperformanceComparisonOfFundamentalESA,BestMorrow_OnTheSignificanceOfCurrentVectorAlignmentInEstablishingTheResonantFrequencyOfSmallSpaceFillingWireAntennas,Best_ElectricallySmallResonantPlanarAntennas}.

Recently, the effect of resonance tuning has once more been taken into account by two different paradigms. The first approach~
\cite{JelinekCapek_OptimalCurrentsOnArbitrarilyShapedSurfaces,2017_TEAT_Gustafsson_Tradeoff_Efficiency_Q} used full-wave treatment of optimal currents on arbitrarily shaped lossy surfaces. The second employed spherical wave expansion~\cite{Pfeiffer_FundamentalEfficiencyLimtisForESA} reaching analytic resonant bounds for spherical geometries.

The purpose of this paper is to show that the effect of resonant tuning on the radiation efficiency of electrically small antennas can be evaluated precisely for arbitrary surface current supports. Furthermore, we demonstrate that for electrically small antennas resonance tuning using realistically lossy materials leads to an unpleasant quartic frequency scaling of dissipation factor.  This opposes the optimistic quadratic frequency scaling predicted by the bounds derived for systems externally tuned by lossless lumped circuits.

This paper is organized as follows. Sections~\ref{assumptions} introduces definitions and restricting assumptions common to the entire manuscript. Section~\ref{intro_example} introduces the radiation efficiency cost of resonance tuning  on a canonical antenna example. Sections~\ref{math} and~\ref{optimization} then show that the introductory observation is of general validity by presenting self-resonant radiation efficiency bounds. The bounds are compared to several realistic designs in Section~\ref{comparison}. Paper is concluded in Section~\ref{concl}.

\section{Assumptions and Definitions}
\label{assumptions}

Here we introduce several definitions and assumptions which help to obtain a mathematically tractable problem:

\begin{enumerate}
	\item{Time-harmonic steady state is assumed with angular frequency $\omega$.}
	\item An antenna is assumed to be tuned to resonance at a given frequency either by antenna current shaping (designed for self-resonance) or by a (potentially lossy) external lumped reactance connected to the antenna terminals.
	\item{An antenna and potential tuning element are made solely of a resistive sheet of given surface resistance $\surfres$. No other material bodies are allowed.}
	\item When particular values of surface resistance $\surfres$ are desired, the skin effect model \mbox{$\surfres = \sqrt{ \left( \omega \mu  \right)/\left( 2\sigma \right)}$} is used with $\mu$ being a permeability and $\sigma$ being a conductivity. This model corresponds~\cite{Jackson_ClassicalElectrodynamics} to a metal sheet of thickness much higher than the penetration depth on which a current flows on one side only.
	\item Within this paper, the radiation efficiency is defined as \mbox{$\radeff = 1/ \left(1 + \dissipfact\right)$} with dissipation factor $\dissipfact$~\cite{Harrington_AntennaExcitationForMaximumGain} being the ratio of cycle mean power lost by heat $\Plost$ to cycle mean power lost by radiation $\Prad$. The power $\Plost$ takes into account conduction losses in the antenna body as well as in the tuning network. By assumptions 3) and 4), conduction losses are the only thermal losses in the system. This definition of radiation efficiency is equivalent to that given in the IEEE Standard~\cite{IEEEStd_antennas} for  metallic antennas, when the matching network made of the same material is considered as a part of the antenna system\footnote{The IEEE Standard Definition~\cite{IEEEStd_antennas} of \Quot{antenna} as \Quot{That part of a transmitting or receiving system that is designed to radiate or to receive electromagnetic waves.} leaves room for interpretation in this regard, particularly for electrically small systems where a radiator and matching or loading elements are constructed of similar materials. Consider for example the equivalence between an inductively base-loaded short monopole and the same monopole tuned by an identical external coil \cite{Harrison1963_MonopoleWithInductiveLoading}.}.
    \item When explicitly needed, the radiation efficiency and dissipation factor counting losses on the untuned antenna structure alone are denoted $\eta_\mathrm{A}$ and $\delta_\mathrm{A}$, respectively.
	\item When an antenna with a well defined input port is tuned to resonance by a series lumped element with impedance \mbox{$Z_\mathrm{T} = R_\mathrm{T} + \J X_\mathrm{T}$}, the dissipation factor of the entire system is given by~\cite{Smith_1977_TAP}
	\BE
	\label{tuning}
	\displaystyle \dissipfact = \dissipfact_{\mathrm{A}} \left( 1 + \frac{\left| \Xin \right|}{\Qtun R_\mathrm{loss}} \right),
	\EE
where \mbox{$Z_\mathrm{A} = R_\mathrm{rad} + R_\mathrm{loss} + \J X_\mathrm{A}$} is the impedance of the antenna, $R_\mathrm{rad}$ and $R_\mathrm{loss}$ distinguish between radiation and ohmic losses \cite{Balanis_Wiley_2005}, and where $\Qtun = \left| X_\mathrm{T} \right| / R_\mathrm{T}$ is the Q-factor of the tuning element.
	\item When applying~(\ref{tuning}), we assume in this paper that small losses inside tuning capacitors can be neglected (\mbox{$\Qtun \to \infty$}), while metallic conductance losses inside tuning inductors must be taken into account.
\end{enumerate}

\section{Introductory Example}
\label{intro_example}

As a motivating example, consider a practical HF band ($3 - 30\,\mathrm{MHz}$) scenario in which an electrically short dipole antenna is tuned to resonance by a series tuning coil. Assume the dipole antenna to be of total length $\ell = 5\,\mathrm{m}$ and made of AWG 6 copper wire ($2.055\,\mathrm{mm}$ radius) \cite{astmb258}. Calculations are carried out from $1\,\mathrm{MHz}$ ($\ell/\lambda = 0.0167$, $ka = 0.0524$) to $28.9\,\mathrm{MHz}$ ($\ell/\lambda = 0.482$, $ka =1.51$), where $k$ is the wavenumber, $\lambda$ is the wavelength and $a \approx \ell/2$ is the radius of the smallest sphere circumscribing the dipole. Note that the highest frequency is selected to be the self-resonance of the antenna, where no tuning inductance is required. The impedance~$Z_\mathrm{A}$ and radiation efficiency~$\eta_\mathrm{A}$ of the dipole antenna alone are calculated using NEC2++ \cite{nec} for the surface resistivity model shown in Section~\ref{assumptions} with conductivity $\sigma = 5.8\cdot10^{7} \, \mathrm{Sm}^{-1}$. The results are shown in Fig.~\ref{fig:dipole}a and compared with an analytical prediction detailed in Appendix~\ref{App0}.
\begin{figure}
	\includegraphics[width=\figwidth cm]{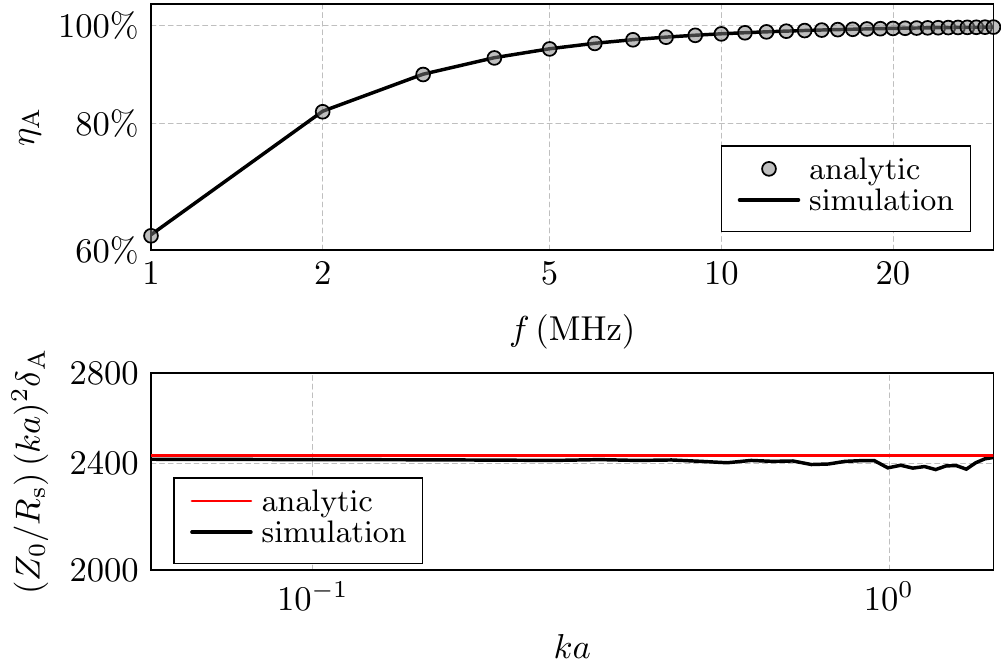}
	\caption{Simulated radiation efficiency $\eta_\mathrm{A}$ (top) and normalized dissipation factor $\delta_\mathrm{A}$ (bottom) of a $5\,\mathrm{m}$ dipole antenna made of copper wire of $2.055\,\mathrm{mm}$ radius.}
	\label{fig:dipole}
\end{figure}

When the dissipation factor corresponding to Fig.~\ref{fig:dipole}a is evaluated and normalized by the surface resistance $R_\mathrm{s}$, it follows the $(ka)^{-2}$ trend expected for electrically small dipole radiators \cite{2012_Stutzman_Antenna_Theory}, as can be seen from Fig.~\ref{fig:dipole}b. Normalization by vacuum impedance \mbox{$Z_0 = \sqrt{\mu_0/\epsilon_0} \approx 120 \pi\,\Omega$} maintains a unitless ordinate.

\begin{figure}
	\includegraphics[width=\figwidth cm]{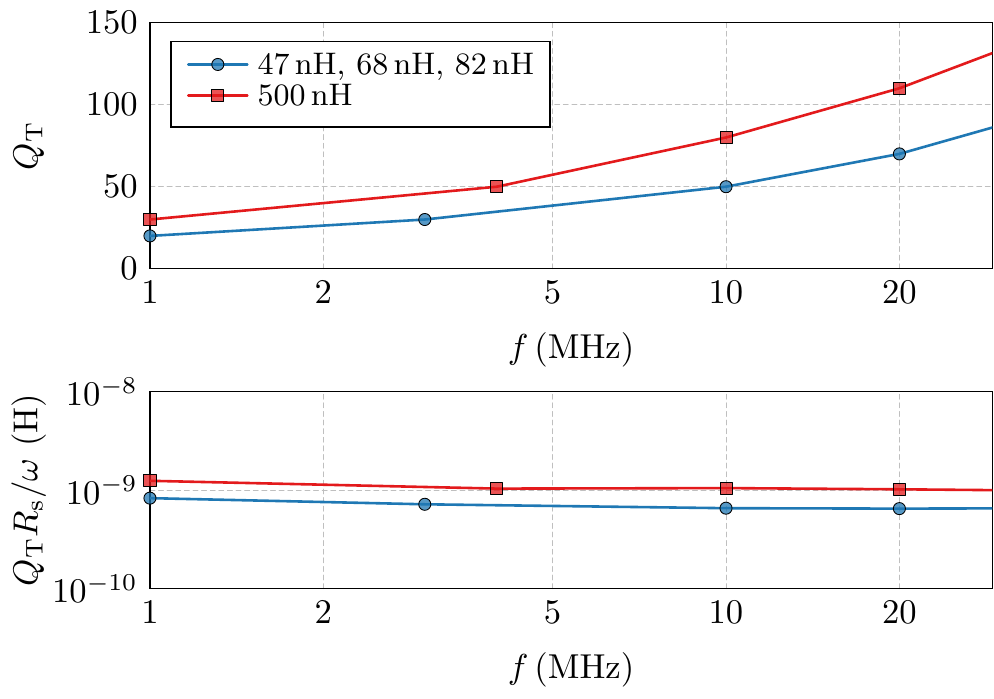}
	\caption{Absolute (top) and normalized (bottom) Q-factors of commercial air-core inductors \cite{coilcraft:square-air-coil}.}
	\label{fig:inductor_q_fit}
\end{figure}

\begin{figure}
	\includegraphics[width=\figwidth cm]{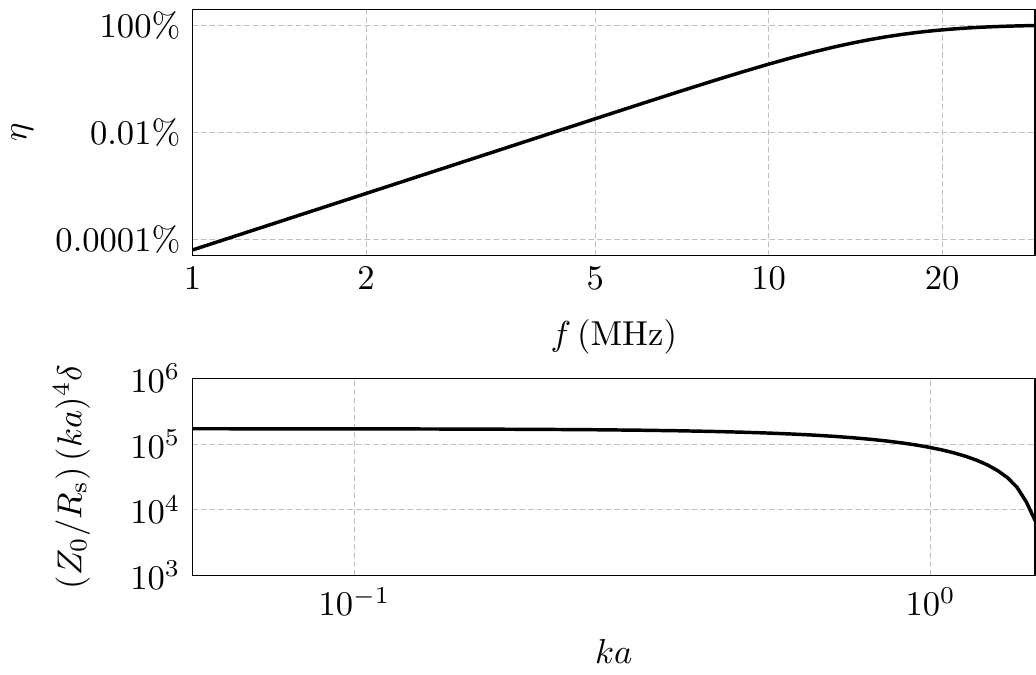}
    \caption{Calculated radiation efficiency (top) and normalized dissipation factor (bottom) of the dipole shown in Fig.~\ref{fig:dipole} tuned by the non-ideal inductors presented in Fig.~\ref{fig:inductor_q_fit}.}
	\label{fig:dipole_total_eff}
\end{figure}

Up until this point, only the properties of the untuned antenna have been considered. To include the dissipation inside the tuning inductor, \mbox{Q-factors $Q_\mathrm{T}$} for commercially available air-coil inductors were obtained from \cite{coilcraft:square-air-coil}. The \mbox{Q-factors} values normalized by the frequency-dependent surface resistance of copper and frequency are shown in Fig.~\ref{fig:inductor_q_fit}. These inductors are thus characterized with $\Qtun \approx 0.7\cdot10^{-9}~\omega / \surfres $ with little dependence on the value of inductance.  This value of $\Qtun$ can thus (for this type of inductor) be substituted to~\eqref{tuning} from which the radiation efficiency of the antenna plus the tuning element can be calculated. The result of this calculation is shown in Fig.~\ref{fig:dipole_total_eff}a and Fig.~\ref{fig:dipole_total_eff}b in absolute and normalized forms.

Comparison of Fig.~\ref{fig:dipole}b and Fig.~\ref{fig:dipole_total_eff}b suggests that when losses in tuning elements are taken into account the following hypotheses are worthy of study:
\begin{itemize}
	\item The radiation efficiency cost of resonance tuning is high, the most important contribution being the lossy tuning element.
	\item Properly normalized dissipation factor of a resonant antenna (self-resonant or tuned) follows a $(ka)^{-4}$ trend.
	\item Dissipation factor normalized as $\left( Z_0 / \surfres \right) \left( ka \right)^4 \dissipfact$ depends\footnote{Radiation efficiency cannot be easily normalized to remove the explicit dependence on size and material parameters. Exceptions are cases when \mbox{$\dissipfact \gg 1$} or \mbox{$\dissipfact \ll 1$.}}, in the electrically small regime, almost exclusively on the shape of an antenna and tuning inductor.
\end{itemize}

Following sections aim to show that the aforementioned observations are valid for cases when antenna and tuning elements are made of arbitrarily shaped lossy surfaces.

\section{Mathematical Tools}
\label{math}

The assumption of an antenna as an arbitrary surface $S$ in~$\mathbb{R}^3$ allows us to employ the electric field integral equation~\cite{Harrington_FieldComputationByMoM}
\BE
\label{EFIE}
- \UV{n} \times \UV{n} \times \left( \V{E}_\T{s} \left( \Jsurf \right) + \V{E}_\T{i} \right) = \surfres \, \Jsurf\quad\text{on $S$,}
\EE
where $\V{E}_\T{s}$ is the scattered electric field, $\V{E}_\T{i}$ the incident electric field, $\Jsurf$ a surface current density, $\surfres$ the surface resistivity, and $\UV{n}$ a unit normal to the surface $S$. For computational purposes the fields and currents tangential to the surface can be modeled as a weighted sum of appropriate basis functions $\{\basisFcn_n\}$, \ie{},
\BE
\label{basis}
\Jsurf \left( \V{r} \right) \approx \sum_n I_n \basisFcn_n \left( \V{r} \right),
\EE
in order to recast \eqref{EFIE} into its matrix form~\cite{Harrington_FieldComputationByMoM}
\BE
\label{RWGEFIE}
\left( \Zmat + \surfres \BFmat \right) \Ivec = \Vvec,
\EE
where $\Ivec$ is a vector of expansion coefficients, $\Vvec$ the excitation vector, \mbox{$\Zmat = \Rmat +  \J \Xmat$} the impedance matrix~\cite{Harrington_FieldComputationByMoM} and $\BFmat$ the Gram matrix~\cite{JelinekCapek_OptimalCurrentsOnArbitrarilyShapedSurfaces}.  Throughout the remainder of this paper, we use \ac{RWG} basis functions~\cite{RaoWiltonGlisson_ElectromagneticScatteringBySurfacesOfArbitraryShape} to represent current densities on simple surfaces, though alternative basis functions may be beneficial or necessary in accurately modeling certain physical or geometrical features (\eg{}, spherical shells, wires, or point contacts).

With the help of the aforementioned matrix formulation the complex power~\cite{Harrington_TimeHarmonicElmagField} can be written as
\BE
\label{eqS21a}
\Prad + \J \Preact \approx \frac{1}{2} \Ivec^\herm \left( \Rmat + \J \Xmat \right) \Ivec,
\EE
and the cycle mean power lost as heat~\cite{Harrington_TimeHarmonicElmagField} can be written as
\BE
\label{eqS23}
\Plost \approx \frac{\surfres}{2} \Ivec^\herm \BFmat \Ivec.
\EE

\section{Optimal Current Maximizing Radiation Efficiency}
\label{optimization}
The classical procedure to find the current distribution on the surface $S$ which maximizes radiation efficiency is to solve~\cite{UzsokySolymar_TheoryOfSuperDirectiveLinearArrays,Harrington_AntennaExcitationForMaximumGain} 
\BE
\label{etaopt01}
\BFmat \Ivec_n = \frac{\delta_n}{\surfres} \Rmat \Ivec_n,
\EE
where matrices $\BFmat$ and $\Rmat$ represent the antenna only, and take the current corresponding to its lowest eigenvalue. The resulting current distribution minimizes normalized dissipation factor~$\dissipfact_\mathrm{A} / \surfres$ and thus maximizes radiation efficiency~$\radeff_\mathrm{A}$. 

\begin{figure}[t]
	\begin{center}
	\includegraphics[width=\figwidth cm]{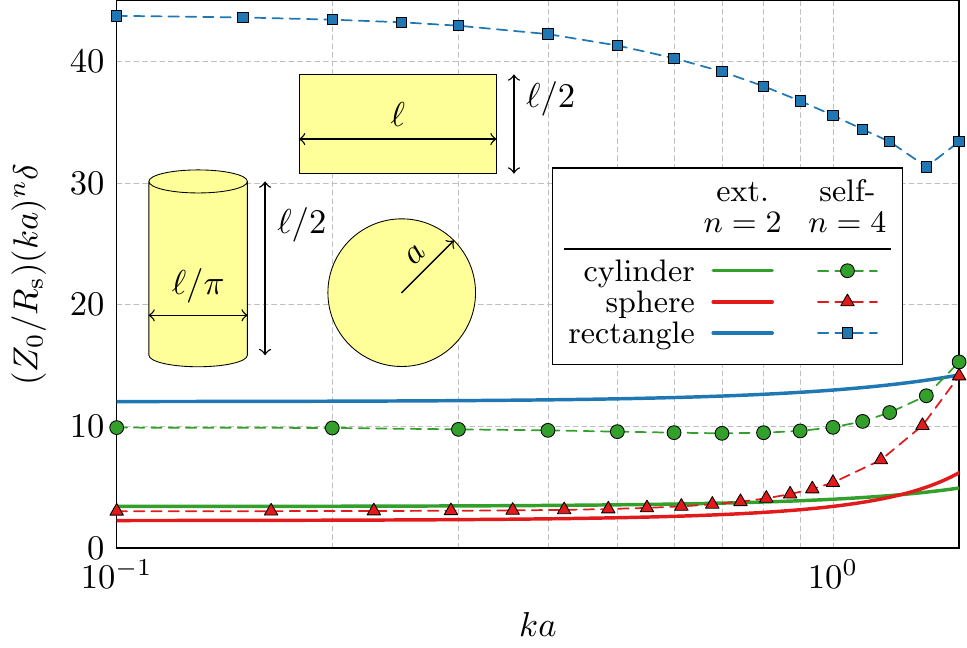}
	\caption{Bound on dissipation factor for selected shapes (cylinder, sphere, rectangle; depicted in insets) of the current carrying region. Solid lines correspond to the external tuning by a lumped lossless element \mbox{(``ext.'')}, while dashed lines correspond to self-resonant bounds \mbox{(``self-'')}.}
	\label{fig1}
	\end{center}
\end{figure}

The solution to \eqref{etaopt01} is not necessarily self-resonant.  If resonance is required, the dissipation factor obtained in \eqref{etaopt01} can only be achieved if there exists a lossless lumped element ($Q_\mathrm{T} \to \infty$) that can tune the current to resonance without affecting dissipation \cite{Fujita_max_gain_APS_2013,2013_Karlsson_PIER,Fujita_max_gain_IEICE_TE_2015,2018_Shahpari_TAP,2012_Stutzman_Antenna_Theory,Fujimoto_Morishita_ModernSmallAntennas,VolakisChenFujimoto_SmallAntennas}. On planar regions this method generates constant current density which is an analytic solution to radiation efficiency maximization in the $ka\rightarrow 0$ limit~\cite{2018_Shahpari_TAP}. Otherwise, the method generally tries to make the current distribution as uniform as possible. This solution neglects the effect of resonance tuning and is depicted for several canonical shapes in Fig.~\ref{fig1} as a function of electrical size~$ka$. Note that these curves scale as $(ka)^{-2}$ for $ka<1$. 

The additional constraint on self-resonance can be incorporated as~\cite{JelinekCapek_OptimalCurrentsOnArbitrarilyShapedSurfaces,2017_TEAT_Gustafsson_Tradeoff_Efficiency_Q}
\begin{equation}
\begin{aligned}
& \T{minimize} && \Ivec^\herm \BFmat \Ivec, \\
& \T{subject~to} &&  \Ivec^\herm \Rmat \Ivec = 1, \\
& && \Ivec^\herm \Xmat \Ivec = 0.
\end{aligned}
\label{eq:S31}
\end{equation}
This optimization problem directly yields the normalized dissipation factor  $\dissipfact / \surfres$. As shown in Appendix~\ref{App1}, its global optimum can be found in a deterministic way. Sample code in~\cite{optimal_dissipation_factor_FileExchange} shows a possible implementation of this procedure and Appendix~\ref{App2} shows convergence of the results for increasing number of discretization elements.

The results generated by (\ref{eq:S31}) are depicted by dashed lines in Fig.~\ref{fig1} for the same problems previously considered. The difference between solid and dashed curves in Fig.~\ref{fig1} at small electrical sizes shows the radiation efficiency cost of resonance tuning. The tuning cost is most easily described by a change from~$(ka)^{-2}$ frequency scaling to~$(ka)^{-4}$ scaling, which agrees well with the example shown in Section~\ref*{intro_example} and with the findings on a spherical shell~\cite{Pfeiffer_FundamentalEfficiencyLimtisForESA,2017_Losenicky_Comment_Pfeiffer}. Results presented in Fig.~\ref{fig1} show that this phenomenon is of general nature for many electrically small objects.

\begin{figure}[t]
	\centering
    \resizebox{4 cm}{!}{\centering \includegraphics[]{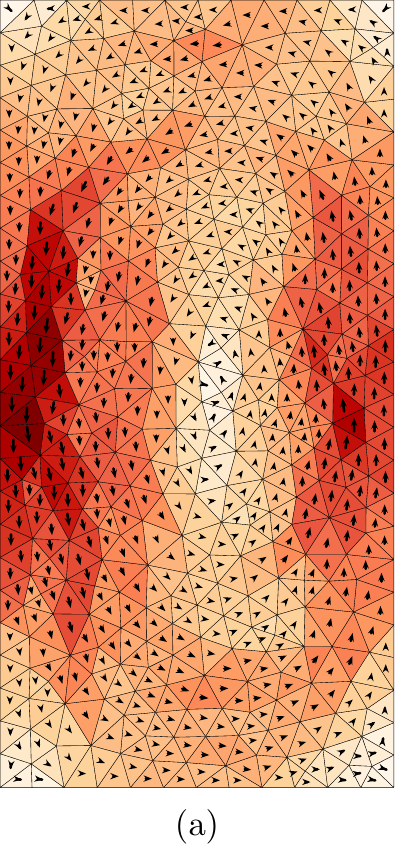}}
	\resizebox{0.3 cm}{!}{\,}	
	\resizebox{4 cm}{!}{\centering \includegraphics[]{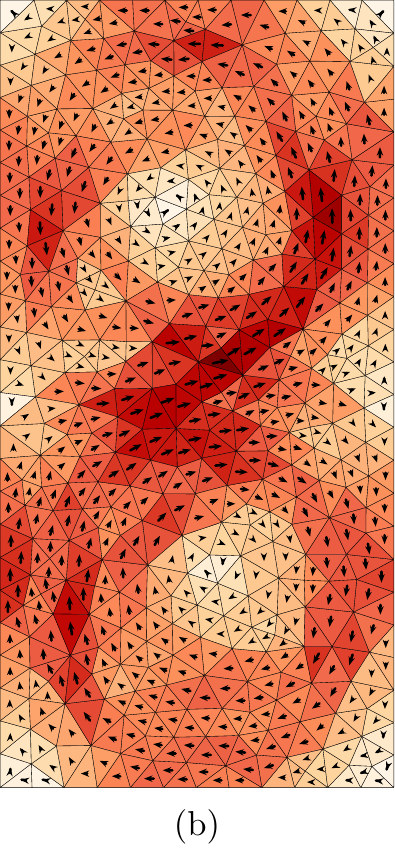}}    
	\caption{Surface current densities minimizing dissipation factor for rectangular region with side ratio~$1$:$2$. The figures correspond to the two lowest minima of \eqref{eq:S31}. Achieved values of normalized dissipation factor are~\mbox{$\left( Z_0 / \surfres \right) \left( ka \right)^4 \dissipfact = 42.7$} and \mbox{$\left( Z_0 / \surfres \right) \left( ka \right)^4 \dissipfact = 58.9$} for the left and right panels, respectively. The used electrical size is~\mbox{$ka = 0.3$}, nevertheless, the current shape is practically unchanged for electrical sizes~\mbox{$ka < 1$}.}
	\label{fig3}
\end{figure}

The self-resonant current maximizing radiation efficiency on a rectangular support (Fig.~\ref{fig1}, rectangular marks) is shown in Fig.~\ref{fig3}a. This current shape is approximately optimal in the full frequency range of Fig.~\ref{fig1} and resembles a combination of electric-dipole-like and magnetic-dipole-like currents as suggested in \cite{Pfeiffer_FundamentalEfficiencyLimtisForESA,2017_Losenicky_Comment_Pfeiffer}. In fact, if the optimal current is evaluated on a spherical shell at small electrical sizes (Fig.~\ref{fig1}, triangular marks), it precisely leads to a resonant combination of TM$_{10}$ and TE$_{10}$ spherical modes \cite{Pfeiffer_FundamentalEfficiencyLimtisForESA,2017_Losenicky_Comment_Pfeiffer}.

\section{Comparison of Bounds with Realistic Designs}
\label{comparison}

\subsection{Self-Resonant Antennas}
\begin{figure}[t]
	\begin{center}
	\centering
	\resizebox{4 cm}{!}{\includegraphics[angle=0]{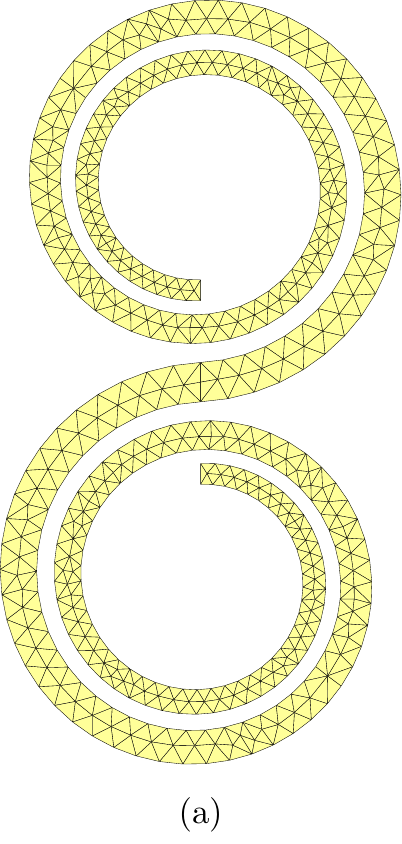}}
	\resizebox{0.3 cm}{!}{\,}	
	\resizebox{4 cm}{!}{\includegraphics[angle=0]{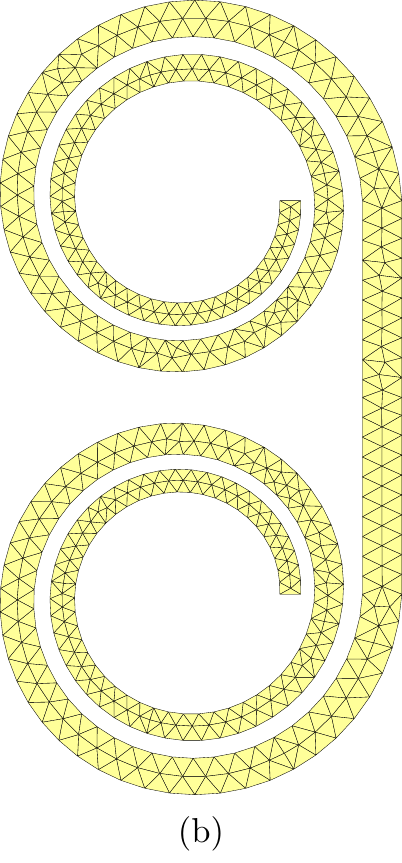}}  
	\caption{Structures mimicking the shape of current density from Fig.~\ref{fig3}b. As in the case of optimal current densities, the meander layout is restricted to a rectangle with edge length ratio approximately $1$:$2$. These particular designs resonate at approximately \mbox{$ka=0.355$} with \mbox{$\left( Z_0 / \surfres \right) \left( ka \right)^4 \dissipfact =158$} (``Julgalt pastry''~\cite{Julgalt}, left panel) or at \mbox{$ka=0.335$} with \mbox{$\left( Z_0 / \surfres \right) \left( ka \right)^4 \dissipfact = 141$} (``Palmier pastry''~\cite{Palmier}, right panel).}
	\label{fig5}
	\end{center}
\end{figure}

\begin{figure}[h]
	\begin{center}
	\includegraphics[width=\figwidth cm]{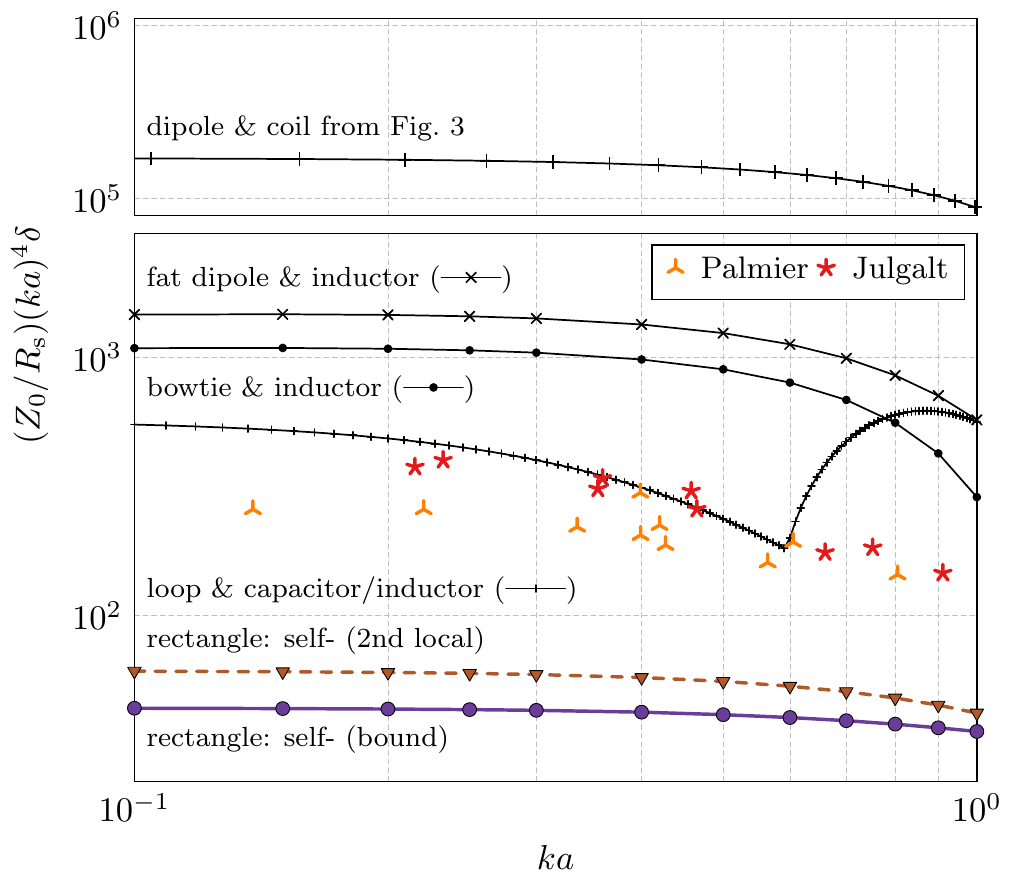}
	\caption{ Comparison of dissipation factors corresponding to the self-resonant bound and the externally tuned bound of a rectangular plate, with dissipation factors of self-resonant spiral meanders from Fig.~\ref{fig5}. Data are also compared to bowtie antenna and fat dipole antenna which were tuned by spiral inductors made of the same resistive sheet as the antenna proper but being ten times smaller in electrical size, \ie{}, \mbox{$ka_\mathrm{T} = ka/10$}. The last comparison is with a rectangular loop antenna tuned by a lossless capacitor below the antiresonance of the loop and by previously mentioned spiral inductor above the antiresonance.}
	\label{fig4}
	\end{center}
\end{figure}

The globally optimal current density depicted in~Fig.~\ref{fig3}a is difficult to realize as a driven antenna current in practice, especially when a design is restricted to have only one localized feed. The method described in~\cite{JelinekCapek_OptimalCurrentsOnArbitrarilyShapedSurfaces}, however, yields also all local optima of the problem in~\eqref{eq:S31}. The local optimum with the second lowest dissipation factor is depicted in Fig.~\ref{fig3}b. The depicted current density suggests that structures from Fig.~\ref{fig5}, which resemble a Julgalt pastry~\cite{Julgalt} and a Palmier pastry~\cite{Palmier}, could be good candidates for approaching the radiation efficiency bound. That this is the case is shown in Fig.~\ref{fig4} although it must be admitted that neither of the structures approach the bound closely (having dissipation factor at least six times higher than the bound).

Though the designs presented here are not necessarily the optimal antenna geometries for attaining maximum radiation efficiency, it has been shown in \cite[design PMD2]{BestMorrow_OnTheSignificanceOfCurrentVectorAlignmentInEstablishingTheResonantFrequencyOfSmallSpaceFillingWireAntennas} that these designs have the highest radiation efficiency among planar meander designs. Despite of the potential suboptimality, both designs follow the $(ka)^{-4}$ trend predicted for self-resonant radiation efficiency bounds.

\subsection{Antennas Externally Tuned by Realistic Components}
\label{sec:exttuned}
\begin{figure}[t]
	\begin{center}
		\includegraphics[width=\figwidth cm]{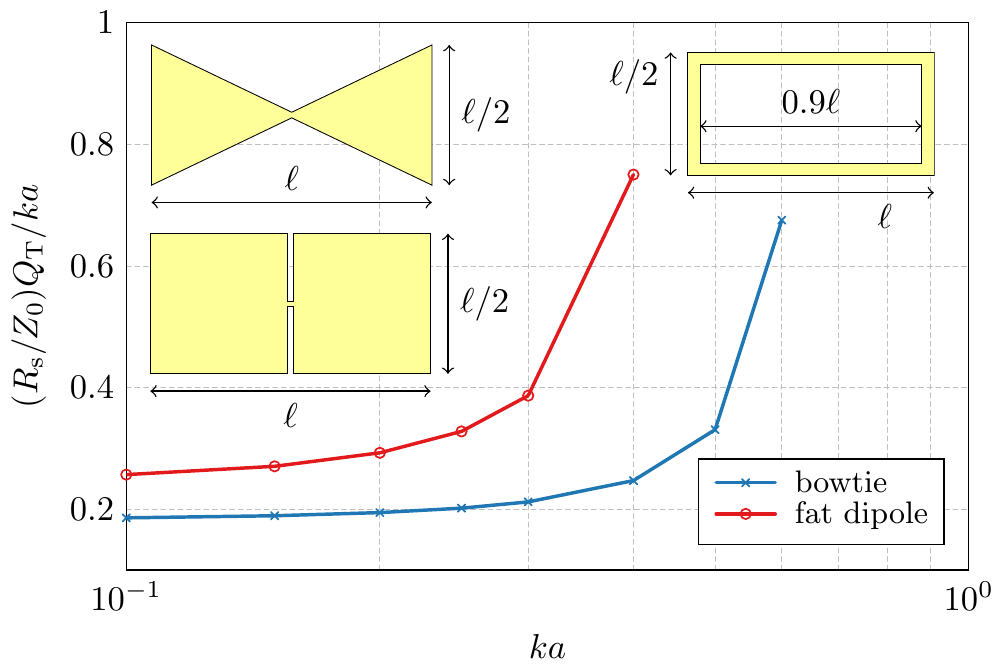}
		\caption{\mbox{Q-factor} of a tuning lumped inductor with which the corresponding antenna will exhibit the same efficiency as the self-resonant bound for a rectangular support of the same size. The curve for a rectangular loop does not exist since the dissipation factor $\dissipfact_{\mathrm{A}}$ of the loop alone is already higher than the self-resonant bound.}
		\label{figQLneeded}
	\end{center}
\end{figure}

Next, we will extend the analysis from Section~\ref{intro_example} and show that antennas externally tuned by realistic components do not surpass the self-resonant bound and, in fact, stay well above the self-resonant spiral meanders when restricted to the same rectangular support. A fat dipole antenna, a bowtie antenna, and a rectangular loop antenna will be used as particular designs, see insets in Fig.~\ref{figQLneeded}. The first two examples are chosen for their space-filling properties, being inspired by the uniform current predicted by \eqref{etaopt01}. The loop is chosen for possibility to use lumped capacitor, \ie{}, low-loss component, at small electrical sizes to achieve resonance.

Let us deal first with electric dipole antennas, a fat dipole and a bowtie, which will be tuned to resonance by realistic lossy inductors. To judge their radiation efficiency performance, imagine that one would desire the  dissipation factor~\eqref{tuning} of an externally tuned antenna to be that of the self-resonant bound from Fig.~\ref{fig4} (circular marks). The relation in~\eqref{tuning} can then be used to extract the \mbox{Q-factor} of the inductor~$Q_\mathrm{T}$ that would be necessary to achieve this goal. The resulting required inductor \mbox{Q-factors}~$Q_\mathrm{T}$ for the fat dipole and bowtie antennas are depicted in Fig.~\ref{figQLneeded}.

\begin{figure}[t]
	\begin{center}
	\includegraphics[width=\figwidth cm]{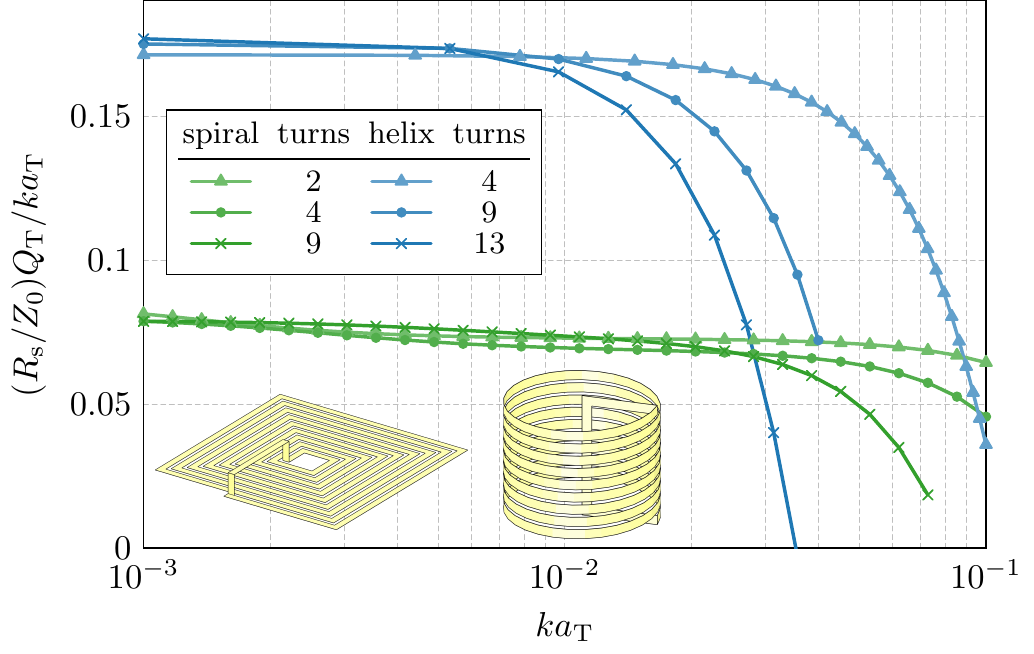}
	\caption{\mbox{Q-factors} of planar spiral inductors and helical inductors for varying numbers of turns. Inductors were made with a spacing between strips equal to one half and one third of the strip width for the spiral and helical geometry, respectively. The diameter of the helical inductor has been made slightly bigger than the inductor length, in order to achieve higher Q-factor \cite{nagaoka1909inductance}. The radius $a_\mathrm{T}$ describes a sphere circumscribing the inductor.}
	\label{figQL}
	\end{center}
\end{figure}

The question now stands if this required inductor \mbox{Q-factor} is achievable by realizable inductors. The negative answer is supported by an example of a planar spiral inductor and a helical inductor, which \mbox{Q-factors} are depicted in Fig.~\ref{figQL}. In both cases, the inductors are made of the same material as the antenna, but due to the used normalization the particular choice of the material is of no relevance. To approximate the assumption of lumped tuning, let us suppose that the inductors are at least ten times smaller\footnote{This may not be possible at very small electrical sizes, since the high capacitive reactance of the radiator will demand inductors of large area.} in electrical size than the antenna, \ie{}, \mbox{$ka_\mathrm{T} \le ka/10$}. In this case the required tuning \mbox{Q-factors} from Fig.~\ref{figQLneeded} are at least an order of magnitude higher than the realizable \mbox{Q-factors} from Fig.~\ref{figQL} and it is important to stress that this conclusion is just weakly dependent on electrical size and is independent on the inductor material. Note here that when for \mbox{$ka_\mathrm{T} = ka/N$}, the values from Fig.~\ref{figQLneeded} can directly be compared to values from Fig.~\ref{figQL} divided by $N$.

For comparison, the bowtie and fat dipole antennas tuned by planar spiral inductors are shown in Fig.~\ref{fig4}. It can be seen that both externally tuned antennas have dissipation factors well above those of the self-resonant spiral meanders.

The case of externally tunned loop antenna differs from electric dipole antennas treated previously. Below the antiresonance of the loop the tuning can be done by low-loss capacitor (here we assume $\Qtun \to \infty$) while above the antiresonance the tuning can be done in the same way as for the bowtie or fat-dipole antenna. The result of this tuning procedure is shown in Fig.~\ref{fig4}. For the particular geometry used, the result is such that dissipation factor $\dissipfact_{\mathrm{A}}$ of the loop alone is already significantly higher than the self-resonant bound for rectangular support and there is thus no possibility to find a tuning network (not even lossless) with which the antenna will reach the radiation efficiency bound. The reason behind this result is that electrically small loop-like current exhibits dissipation factor that already scales with frequency as $(ka)^{-4}$ and the tuning network can only worsen this behavior. In other words, it can be stated that a high cost of resonance tuning in all electrically small radiators is presented by the loop-like current that is used for resonance tuning.

One may ask whether using physically large inductors or superconducting materials in the matching network is a way around the dissipation factor scaling with $(ka)^{-4}$.  If matching components are made larger while still being considered approximately lumped (\eg{}, \mbox{$ka_\mathrm{T} = ka/8$} rather than \mbox{$ka_\mathrm{T} = ka/10$} in the preceding analysis), then the above analysis holds, though the precise numerical results may slightly change.  If, however, the tuning network is made similar in size to the antenna itself, then it becomes appropriate to include the spatial support of the tuning network into the derivation of the radiation efficiency bounds.  Hence, the tuning network becomes part of the current optimization problem in \eqref{eq:S31} where it may even be leveraged as a source of radiation. In any case the dissipation factor of the system will scale as $(ka)^{-4}$.

\section{Conclusion}
\label{concl}

It has been shown that radiation efficiency bounds assuming external tuning by lossless lumped elements are overly optimistic and that tighter self-resonant bounds can easily be calculated. It was further demonstrated that selected self-resonant antennas can approach this bound and that their radiation efficiency surpasses that of the non-resonant antennas tuned by realistic reactances. An important conclusion is that when resonance tuning is demanded, an unpleasant frequency scaling of dissipation factor $\left(ka\right)^{-4}$ must be assumed for electrically small antennas, rather than the previously predicted $\left(ka\right)^{-2}$ scaling.


\appendices
\section{}
\label{App0}
An analytical model used in Section~\ref{intro_example} assumes a thin cylindrical dipole with length $\ell$, radius $r_\mathrm{w}$ and current distribution
\begin{equation}
\label{App:dipole:I}
I\left( \zeta \right) = {I_0} \sin \left( {k\left( {\frac{\ell}{2} - \left| \zeta \right|} \right)} \right),
\end{equation}
where $\zeta \in \left[-\ell/2,\ell/2\right]$. Radiation resistance~$R_\mathrm{rad}$ of such a current reads~\cite{Balanis_Wiley_2005}
\begin{equation}
\label{App:dipole:Rrad}
R_\mathrm{rad} = \frac{Z_0}{\pi}\displaystyle\int\limits_0^1 \displaystyle \frac{\left( \displaystyle \cos \left( \frac{k\ell}{2} \right) - \cos \left( \frac{k\ell}{2}\varsigma \right) \right)^2}{\displaystyle \sin^2\left( \frac{k\ell}{2} \right)\left( {1 - \varsigma^2} \right)} \, \mathrm{d}\varsigma.
\end{equation}
Assuming the dipole made of a cylindrical surface (without end caps), which is covered by surface resistance~$R_\mathrm{s}$, the loss resistance~$R_\mathrm{loss}$~\cite{Balanis_Wiley_2005} can be calculated as
\begin{equation}
\label{App:dipole:Rloss}
\frac{R_\mathrm{loss}}{R_\mathrm{s}} = \left( \frac{\ell}{ 4\pi r_\mathrm{w} } \right) \frac{\displaystyle 1 - \frac{\sin \left( {k\ell} \right)}{k\ell}}{ \displaystyle \sin^2\left( {\frac{k\ell}{2}} \right)}.
\end{equation}
The ratio of the loss resistance~$R_\mathrm{loss}$ and the radiation resistance~$R_\mathrm{rad}$ is the untuned dissipation factor.  Expanding and taking the leading order term of the quotient of \eqref{App:dipole:Rloss} and \eqref{App:dipole:Rrad} yields
\begin{equation}
\label{App:dipole:deltaA}
{\left( \frac{k\ell}{2} \right)^2} \frac{Z_0}{R_\mathrm{s}}\delta_\mathrm{A} \approx \displaystyle \frac{\ell}{r_\mathrm{w}}.
\end{equation}
This agrees with the short dipole approximation of triangular current distribution \cite{2012_Stutzman_Antenna_Theory} and demonstrates explicitly that, at small electrical size, the untuned dissipation factor scales as $\left( k\ell \right)^{-2}$, which is graphically presented in Fig.~\ref{fig:dipole}.
Turning now to the tuned dissipation factor, formula~\eqref{tuning} can be rewritten as
\begin{equation}
\label{App:dipole:delta}
\delta  = \delta_\mathrm{A} \left( 1 + \frac{ \displaystyle k\ell\left| X_\mathrm{A} \right| }{\left( \displaystyle k\ell \right)^2 \left( \displaystyle \frac{Q_\mathrm{T} R_\mathrm{s}}{k\ell} \right)\left( \displaystyle \frac{R_\mathrm{loss}}{R_\mathrm{s}} \right)} \right).
\end{equation}
At small electrical sizes $ka \to 0$ the input reactance of a wire dipole is capacitive $X_\mathrm{A} \propto k^{-1}$ and the term ${k\ell\left| {{X_{\mathrm{A}}}} \right|}$ is independent of frequency. The same holds for the terms $Q_\mathrm{T} R_\mathrm{s} / \left(k\ell\right)$ and $ R_\mathrm{loss} / R_\mathrm{s}$ in~\eqref{App:dipole:delta} as can be seen from Fig.~\ref{fig:inductor_q_fit} and~\eqref{App:dipole:Rloss}. This explicitly shows how the resonance tuning is responsible for the change from  $\left( k\ell \right)^{-2}$ scaling to  $\left( k\ell \right)^{-4}$ scaling when electrical size is small.

\section{}
\label{App1}
This appendix briefly describes the method used to solve optimization problem~(\ref{eq:S31}). See \cite{optimal_dissipation_factor_FileExchange} for an example of MATLAB implementation.

The solution is approached by a dual formulation \cite{NocedalWright_NumericalOptimization} in which one maximizes so-called dual function
\BE
\label{Eq:App:gfunc}
g\left( \lambda _1,\lambda _2 \right) = \mathop{\inf }\limits_{\Ivec} \big[  \mathcal{L} \left( \Ivec,\lambda _1,\lambda _2 \right) \big],
\EE
where
\BE
\label{Eq:App:lagrangian}
\mathcal{L} \left( \Ivec,\lambda _1,\lambda _2 \right) = \Ivec^{\mathrm{H}} \BFmat \Ivec - \lambda _1 \Ivec^{\mathrm{H}} \Xmat \Ivec - \lambda _2 \big( \Ivec^{\mathrm{H}} \Rmat \Ivec - 1 \big)
\EE
is the Lagrangian corresponding to~(\ref{eq:S31}) and $\mathop{\inf }$ denotes infimum. The supremum of dual function $g\left( \lambda _1,\lambda _2 \right)$ is the lower bound \cite{NocedalWright_NumericalOptimization} to the original problem~(\ref{eq:S31}). In~\cite{CapekGustafssonSchab_MinimizationOfAntennaQualityFactor} it was however shown that for radiation problems of this kind there is no dual gap \cite{NocedalWright_NumericalOptimization} present and the supremum of $g\left( \lambda _1,\lambda _2 \right)$ is the global optimum of~(\ref{eq:S31}). Since function $g\left( \lambda _1,\lambda _2 \right)$ is concave~\cite{NocedalWright_NumericalOptimization} it is assured that this global optimum can be approached to an arbitrary precision in a finite number of steps.

In the code available at~\cite{optimal_dissipation_factor_FileExchange}, the supremum of $ g\left( \lambda _1,\lambda _2 \right) $ is searched only among stationary points of the Lagrangian~(\ref{Eq:App:lagrangian}), which are guided by
\BE
\left( \BFmat - \lambda _1 \Xmat \right) \Ivec = \lambda _2 \Rmat \Ivec.
\EE
This tremendously narrows the solution space and fixes the relation between Lagrange multipliers $\lambda _1$ and $\lambda _2$. The dual function~(\ref{Eq:App:gfunc}) is therefore a function of single variable ($\lambda _1$ in the code~\cite{optimal_dissipation_factor_FileExchange}) and its maximum can be obtained, for example, by a golden-section search~\cite{golden_section_search}.

To further ease the computational burden in the code~\cite{optimal_dissipation_factor_FileExchange} and following~\cite{JelinekCapek_OptimalCurrentsOnArbitrarilyShapedSurfaces}, the original problem~(\ref{eq:S31}) is further projected onto macrobasis functions generated by the following eigenvalue problem
\BE
\Xmat \Ivec_n = \nu_n \BFmat \Ivec_n.
\EE
This macrobasis is favorable in diagonalizing two of the three underlying operators. The unknown current vectors $\Ivec$ and operators $\BFmat$, $\Xmat$, $\Rmat$ in the \ac{RWG} basis are projected into this new macrobasis. This eases the computational burden since, to a high degree of precision, the optimal solution is composed of a few macrobasis functions~\cite{JelinekCapek_OptimalCurrentsOnArbitrarilyShapedSurfaces}.

\section{}
\label{App2}
\begin{figure}[h]
	\includegraphics[width=\figwidth cm]{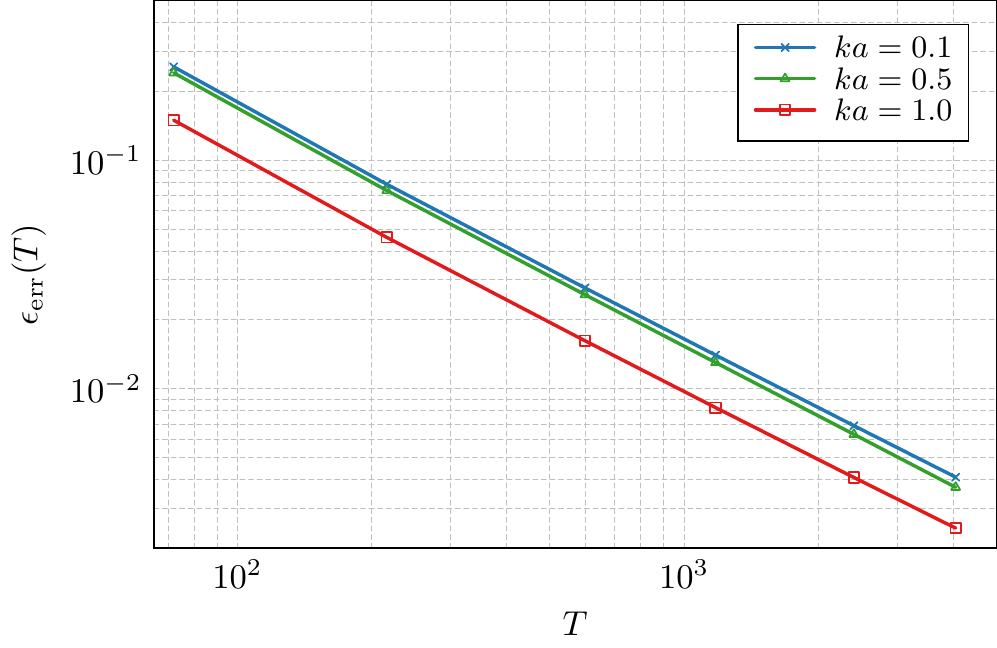}
	\caption{Relative error of a self-resonant dissipation bound corresponding to a spherical shell of radius $a$ evaluated by the method detailed in Appendix~\ref{App1} with respect to an analytical result presented in~\cite{2017_Losenicky_Comment_Pfeiffer}. Numerical results were calculated in AToM~\cite{atom}.}
	\label{fig:convergence}
\end{figure}

The lower bound on a self-resonant dissipation factor~$\delta$ corresponding to a current flowing on a spherical shell is known analytically~\cite{2017_Losenicky_Comment_Pfeiffer}. The spherical shell thus provides ideal grounds for testing the convergence of the optimization scheme outlined in Appendix~\ref{App1} which was used to generate results in Fig.~\ref{fig1}. To that point, several discretizations of a spherical shell have been used and the resulting self-tuned dissipation factors have been compared with the above mentioned analytical result. The relative error~$\epsilon_\mathrm{err}\left(T\right)$ of the numerical evaluation is shown in Fig.~\ref{fig:convergence} for a particular choice of $T = \left\{72, 216, 600, 1176, 2400, 4056\right\}$ triangles.

It can be observed that the optimization scheme presented in Appendix~\ref{App1} is numerically robust, achieving precision gain of approximately one digit per one order in number of triangles~$T$. Although the analytical data for other shapes are not available, it can be expected that the precision for other canonical shapes presented in Fig.~\ref{fig1} will be similar, provided that their geometries and current paths are well represented by the chosen basis.



\end{document}